\documentstyle[12pt,epsf]{article}

\def\dfrac{\displaystyle\frac}
\newcommand{\bea}{\begin{eqnarray}}
\newcommand{\eea}{\end{eqnarray}}
\newcommand{\be}{\begin{equation}}
\newcommand{\ee}{\end{equation}}
\newcommand{\cL}{{\cal L}}
\newcommand{\ve}{{\varepsilon}}

\def\ut#1{\mathop{\vtop{\ialign{##\crcr
     $\hfil\displaystyle{#1}\hfil$\crcr\noalign
     {\kern1pt\nointerlineskip}\hbox{$\hfil\sim\hfil$}\crcr
     \noalign{\kern1pt}}}}}
\def\undersim{\ut}

\def\fun#1#2{\lower3.6pt\vbox{\baselineskip0pt\lineskip.9pt

\ialign{$\mathsurround=0pt#1\hfill##\hfil$\crcr#2\crcr\sim\crcr}}}
\newcommand{\ap}[3]{{\it  Ann. Phys. }{{\bf #1} {(#2)} {#3}}}
\newcommand{\app}[3]{{\it  Astropart. Phys. }{{\bf #1} {(#2)} {#3}}}
\newcommand{\hpa}[3]{{\it  Hel. Phys. Acta }{{\bf #1} {(#2)} {#3}}}
\newcommand{\mpl}[3]{{\it  Mod. Phys. Lett. }{{\bf #1} {(#2)} {#3}}}
\newcommand{\lnc}[3]{{\it  Lett. Nuov. Cim. }{{\bf #1} {(#2)} {#3}}}
\newcommand{\np}[3]{{\it  Nucl. Phys. }{{\bf #1} {(#2)} {#3}}}
\newcommand{\pr}[3]{{\it Phys. Rev.}{{ \bf #1} {(#2)} {#3}}}

\newcommand{\prl}[3]{ {\it Phys. Rev. Lett.}{{ \bf #1} {(#2)} {#3}}}
\newcommand{\pl}[3]{{\it  Phys. Lett. }{{\bf #1} {(#2)} {#3}}}

\newcommand{\ptp}[3]{{\it  Prog. Theor. Phys.}{{\bf #1} {(#2)} {#3}}}

\begin{document}
\begin{titlepage}

\begin{flushright}
UUITP-2/95\\ March 1995\\
hep-ph/9503459
\end{flushright}
\vspace{1 cm}

\begin{center}
{\huge The Polarization of a QED Plasma in a Strong Magnetic Field}
\end{center}

\vspace{1 cm}

\begin{center}
{\large Ulf H. Danielsson and Dario Grasso} \\
\vspace{.7 cm}
Department of Theoretical Physics, Uppsala University\\ Box 803, S-751
08 Uppsala, Sweden
\end{center}

\vspace{1 cm}

\centerline{{\Large\bf Abstract}}

\normalsize

\begin{quote}
In this paper we study the polarization tensor of photons in a QED
plasma in the presence of a magnetic field. We do it both at vanishing
and at finite temperature. We use two different methods to compute the
polarization tensor components $\Pi^{\mu\nu}$ within the one-loop
approximation.  The first starts from the effective Lagrangian of the
system and relates $\Pi^{\mu\nu}$ to the thermodynamic quantities. The
second makes use of the electron propagator in an external magnetic
field. In this second approach we use an imaginary time
formalism. These methods give consistent results in the first
non-vanishing order in the photon 4-momentum.  Beyond this limit the
first method is not applicable and the introduction of the second can
not be avoided. We give some physical interpretations of our results.
\end{quote}
\vspace{1cm}
PACS numbers: 12.20Ds., 11.10.W, 05.30.Fk.
\end{titlepage}

\newpage

\section{Introduction}

The investigation of the electromagnetic properties of a relativistic
plasma in a strong external magnetic field is interesting both from a
theoretical point of view and for possible applications in cosmology
and astrophysics. Large magnetic fields are known to be present in
neutron stars \cite{neustar}, where $B \sim 10^{12}$ G, in supernovas
\cite{supernova}, where $B \sim 10^{14}$ G, and several models foresee
very large magnetic fields in the early Universe \cite{Kronberg}.
When such strong magnetic fields are present the propagation of
electromagnetic waves in a plasma is considerably modified with
respect to the free field case.  These strong fields can be not
treated like a perturbation. Indeed, the tensorial structure of the
polarization tensor need to be completely reconsidered \cite{Rojas}.

According to the pioneering approach of Fradkin \cite{Fradkin} one of
our aims is to relate the electromagnetic properties of a QED plasma
in a strong magnetic field to the thermodynamic properties of the
system. This should give the reader a physically more transparent
interpretation of the results that partially are present already in
the literature, as well as of the new results that we obtain here.

All the thermodynamic properties of a QED plasma can be derived
starting from the effective Lagrangian of the system.  Assuming ${\bf
B(x)} = (0,0,B)$ the Lagrangian can be written
\be
\cL^{\rm eff} = \cL^{\rm eff}(B) + \cL^{\rm eff}(B,T,\mu)        ,
\ee
where the vacuum contribution at the one loop level is
\cite{Schwinger}
\be
 \cL^{{\rm eff}} (B)=-\frac{1}{8\pi^2} \int_0^\infty \frac{ds}{s^3}
 \left[ eBs \coth(eBs)-1-\frac13 (eBs)^2 \right] \exp(-m^2s) ,
\ee
whereas the one-loop matter contribution is
\bea
\lefteqn{\cL^{\rm eff}(B,\mu,T) = \frac {\ln Z}{\beta V}=}
\nonumber \\
&&\frac 1 {\beta} \frac{|e|B}{2\pi^2}
\sum_{n=\hat 0}^\infty
\int_{-\infty}^\infty d\!k_{3} \left\{
\ln[ 1+e^{-\beta(E_n(k_{3})-\mu)}] +
\ln[ 1+e^{-\beta(E_n(k_{3})+\mu)}] \right\}
\label{lbt}.
\eea
Here
\be
E_n(k_{3}) = \sqrt{k_{3}^2 + 2neB + m^2}
\ee
is the energy of the $n$-th Landau level at tree level.  The sum
$\sum_{n=\hat 0} \equiv {1 \over 2 } \sum_n (2 -
\delta_{n0})$
takes into account the double spin degeneration of the Landau levels
with the exception of the lowest.  Equation (\ref{lbt}) follows from a
simple phase space consideration {\footnote{ Although Eq.(\ref{lbt})
can be obtained from the fermion propagator as done in
refs.\cite{Dittrich}\cite{Elmfors93}, we prefer to adopt here this
simpler physical interpretation.}}.  In fact, due to the coalescing of
the transverse momentum states into those of a two-dimensional
harmonic oscillator, the number of available states for any given
value of $n$ is
\be
\frac {V|e|B}{(2\pi)^2}(2 - \delta_{n0})\,dk_{3}~~~.
\label{phasesp}
\ee

One of the most important effects of the modification of the electron
phase-space is the ``pair generation" that takes place when magnetic
fields larger than $B_c \equiv m^2/e$ are applied to the plasma. We
use quotation marks in order not to confuse the reader about the
meaning of pair generation in this context.  Clearly, a constant
magnetic field can not induce pair generation from the
vacuum. However, things can be different at finite temperature and/or
density.  Indeed, the number density of electrons+positrons is
\cite{Miller}
\be
n = \frac {|e|B}{2\pi^2} \sum_{n=\hat 0}^\infty \int_{-\infty}^\infty
        dk_{3} \left\{ \frac 1 {1+e^{\beta (E_n(k_{3})-\mu)}} + \frac
        1 {1+e^{\beta(E_n(k_{3})+\mu)}} \right\}~~~.  \label{npairs}
\ee
It increases roughly linearly with B when $B \gg B_c$ and $T
\undersim{<} (|e|B)^{1/2}$. This phenomenon can be understood as a
shift in the equilibrium of photons and pairs \cite{Dittrich}: the
equilibrium of the process $e^+e^- \leftrightarrow \gamma$ moves to
the left owing to the growing of the number of available states that
electrons and positrons can occupy in the lowest Landau level.  Thanks
to the amplification of its phase-space this level is practically the
only occupied level when $eB \gg T^2$. Since $B$ modifies the density
of the charge carriers, it is reasonable to expect that the
electromagnetic properties of the plasma have to be affected by strong
magnetic fields.  The charge neutrality of the plasma is preserved,
provided that $\mu = 0$, since the increasing of the electron and
positron energy densities balance each other. Indeed, the charge
density is
\bea
\lefteqn{\rho(\mu) = \frac {\partial \cL^{\rm eff}(B,T,\mu)}
{\partial \mu} = }\\ && \frac {|e|B}{2\pi^2} \sum_{n=\hat 0}^\infty
\int_{-\infty}^\infty dk_{3} \left\{ \frac 1 {1+e^{\beta
(E_n(k_{3})-\mu)}} - \frac 1 {1+e^{\beta(E_n(k_{3})+\mu)}} \right\}
\label{pairs}\ .\nonumber
\eea

The magnetization plays an essential role in determining the screening
properties of the plasma.  Like the energy density, the magnetization
${\bf M}$ can be obtained starting from the effective Lagrangian
\be
{\bf M} = \frac {\partial \cL}{\partial {\bf B}} =
\frac {\partial \cL^{\rm eff}(B)}{\partial {\bf B}} +
\frac {\partial \cL^{\rm eff}(B,T,\mu)}{\partial {\bf B}}~~~.
\ee
For example, the matter contribution to the magnetization in the limit
$T = 0$ is (we refer the reader to ref. \cite{Elmfors93} for a more
general expression) $$ M(B,0,\mu) = $$
\be
{|e| \over 2 \pi ^{2}}
\sum _{n=\hat{0}}^{\left[ {\mu ^{2}-m^{2} \over 2eB} \right]}
\left( \mu \sqrt{\mu ^{2} -m^{2}-2eBn} -
 (m^{2} +4eBn) \log {\mu + \sqrt{\mu ^{2} -m^{2} -x} \over
\sqrt{m^{2}+x}} \right) . \label{magn}
\ee
Note that in contrast to what happens for $\rho$, a component of the
magnetization depends on the vacuum part of the effective
Lagrangian. In other words, the vacuum polarizes in strong magnetic
fields. The relative strength of the vacuum and matter polarization
has been evaluated in ref.\cite{Elmfors93}.  The vacuum contribution
to the magnetization exceeds the matter contribution if $eB \gg
T^2,~|\mu^2 - m^2|$.  We do not consider here the other vacuum
contributions to the polarization tensor since they are already widely
reported in the literature (see e.g. ref.  \cite{Dit-Reu}).

In the next section we will show that the components of the
polarization tensor can be related to the thermodynamic quantities
$\rho$, ${\bf M}$ and the magnetic susceptibility $\chi = \partial
M/\partial B$. In section 3 we write the general expression of the
polarization tensor in terms of the fermion propagator in an external
magnetic field. In section 4 we apply this expression to get the
components of $\Pi^{\mu\nu}$ in the static limit and verify that they
coincide with the results obtained in section 2. In section 5 we give
examples of computations of some components of $\Pi^{\mu\nu}$ beyond
the static limit.  Finally, section 6 contains our conclusions.

\section{The screening operator in terms of $\cL^{\rm eff}$}

The polarization operator for a QED plasma is defined by
\be
\Pi_{\mu\nu}(x,x') \equiv i~\frac {\delta <j_\mu(x')> }
{\delta A_\nu(x)} = i~\frac {\delta}{\delta A_\nu(x)}\,
\frac {\delta \Gamma^{\rm eff}}{\delta A_\mu(x')}~~~.
\label{def}
\ee
where $\Gamma^{\rm eff} = \int d^4x\,\cL^{\rm eff}$ is the effective
action. If the plasma has a non-vanishing chemical potential the tree
level Lagrangian is
\be
 \cL = -\frac14 F_{\mu\nu}F^{\mu\nu} +\bar{\psi}(i {\partial
 \kern-0.5em/} -e {A\kern-0.5em/} - \gamma_0 \mu -m)\psi~~~.
\label{tree}
\ee
The form of Eq. (\ref{def}) remains unchanged if we replace $A(x)$
with $\tilde A(x) \equiv (A^0 + \frac 1e \mu, {\bf A})$.  Note that
only this combination of the chemical potential and the vector
potential is physically meaningful
\cite{Elmfors94}. In fact, whenever $A_0$ is changed, the
polarization of the plasma rearranges so to fulfil $\mu + eA_0 = {\rm
const.}$ at the equilibrium \cite{Kapusta}.  Thus, $\mu$ is determined
by the charge distribution of the plasma, and its value may depend on
the position.  It is then convenient to work in the gauge $A_0 = 0$ in
order to have a well defined $\tilde A(x) = \mu = const.$ for the
whole plasma.

As a first application of the above formula we compute the limit for
$x \rightarrow x'$ of $\Pi^{00}$. We then have
\bea
\Pi^{00}(B,T,\mu) &=&e^2 \frac {\partial^2 \cL^{\rm
eff}(B,T,\mu)} {\partial \mu^2} = e^2 {\partial^2 \rho\over \partial
\mu^2} =
\label{pi00}\\
&=& {|e|^3B\beta \over (2\pi )^{2}}\sum_{n = \hat{0}}^{\infty}
{\int_{0}^{\infty} dk_3~ {\left( {1 \over \cosh ^{2}\left( {\beta (E_n
(k_{3}) + \mu )
\over 2}\right)}  + {1 \over \cosh ^{2}\left( {\beta (E_n  (k_{3}) -
\mu ) \over 2}
 \right)}\right)}} \nonumber
\eea
where, assuming a uniform ${\bf B}$ and working in the Euclidean
space, we have used $\Gamma^{\rm eff}_{\rm mat} = -i \beta V
\cL (B,T,\mu)$.

 It is worthwhile to compare this result with the corresponding
quantity obtained for $B = 0$ \cite{Midorikawa}
\be
\Pi^{00}(0,T,\mu) =
{e^2\beta \over 2(2\pi)^3}{\int_{0}^{\infty} d^3{\bf k}{\left( {1
\over \cosh ^{2}\left( {\beta (E + \mu ) \over 2}\right)} + {1 \over
\cosh ^{2}\left( {\beta (E - \mu )\over 2} \right)}\right)}}
\ee
Keeping in mind Eq.(\ref{phasesp}) it is evident at a glance that the
effect of the magnetic field on $\Pi^{00}$ is completely mediated by
the modification of the electron phase space. The increase of the
screening power of static electric fields when overcritical magnetic
fields are applied to the plasma can also be interpreted as due to the
increase of the number of pairs (see Eq. (\ref{npairs})) as we
anticipated in the introduction. In this sense the magnetic field
plays a role analogous to that played by the temperature.

Let us now consider the $\Pi^{0i}$ components ($i = 1,2,3$).  In the
same gauge as we used above, we have
\be
\Pi_{0i}(x,x') = i~\frac \delta{\delta \tilde A^0}
\frac \delta {\delta \tilde A^i}~\Gamma^{\rm eff} =
ie\,\ve_{ijk} \partial_k \frac {\delta \rho}{\delta B^j(x')}.
\ee
where we used the chain rule for the functional derivative to get
$\dfrac \delta {\delta \tilde A^i} = \ve_{ij}^{k} \partial_k
\dfrac \delta {\delta B^j}$.
Whereas $\Pi_{03} = 0$ identically and $\Pi_{01}=\Pi_{02} = 0$ if the
momentum insertion is zero or parallel to the field, the transversal
components receive a non-vanishing contribution if the momentum
insertion is perpendicular to the field. More precisely, the Fourier
transform of (14) in terms of the photon momentum components $p_{i}$
is
\be
\Pi_{0i} = ie \ve_{ij}~p_i \frac {\partial\rho}{\partial
B} + O(p^2) \quad i = 1,2 \label{pi0i}
\ee
while $\Pi_{i0} = \Pi_{0i}^*$. These terms are related to the to the
interaction of the net charge of the plasma with the magnetic field
and contribute to the Hall conductivity.  For the spatial components
we get the general formula
\be
\Pi_{ij} = \ve_{i}^{kl}\ve_{j}^{mn}~p_kp_m \left[ \left(\delta_{nl} -
\frac {B_nB_l}{B^2}\right)  \frac 1 B \frac \partial {\partial
B} + \frac {B_nB_l}{B^2} \frac {\partial^2} {\partial B^2}
\right] \cL^{\rm eff}  \label{piij}
\ee
This gives
\bea
\Pi _{33} &=& p_{2}^2 \frac 1 B~M~~~; \label{pizz}\\
\Pi _{11} &=&   p_{3}^2 \frac 1 B~M + p_{2}^2~ \chi~~~;\label{pixx}\\
\Pi _{22} &=&   p_{3}^2 \frac 1 B~M + p_1^2~ \chi~~~;\label{piyy}\\
\Pi_{12} &=&\Pi_{21} =  p_1p_{2}~ \chi~~~;\\
\Pi_{13} &=&\Pi_{31} =  - p_{3}p_1 \frac 1 B~M ~~~;\\
\Pi_{23} &=&\Pi_{32} = p_{3}p_{2} \frac 1 B~M ~~~.\label{piyz}
\eea
It is worthwhile to note the symmetry of the spatial part of the
polarization tensor. This agrees with the tensorial structure obtained
in ref.\cite{Rojas}.

\section{The electron propagator in an external magnetic field}

Although the results of the previous section have been obtained
without any use of the fermion propagator $S$, a treatment addressed
to determine the components of the polarization tensor beyond the
leading order in momenta can not leave $S$ out of consideration.  The
fermion propagator in a constant magnetic field is given by
\cite{Kob-Sak}
\bea
S(x,x') &=& \sum_{n=0}^{\infty}\int {d\omega dk_{2} dk_{3} \over (2\pi
)^{3}} e^{-i\omega (t-t') +ik_{2} (y-y') +ik_{3}(z-z')}
\nonumber\\
&&\times {1 \over \omega ^{2} -k_{3}^{2}-m^{2}-2eBn +i\epsilon}
S(n;\omega ,k_{2},k_{3})
\eea
where $$ S(n;\omega ,k_{2},k_{3}) = $$
\be
\left( \matrix{
mI_{n,n} & 0 & -(\omega + k_{3})I_{n,n} & -i\sqrt{2eBn}I_{n,n-1}\cr 0
 & mI_{n-1,n-1} & i\sqrt{2eBn} I_{n-1,n} & -(\omega -
 k_{3})I_{n-1,n-1} \cr -(\omega - k_{3})I_{n,n} & i\sqrt{2eBn}
 I_{n,n-1} & mI_{n,n} & 0 \cr -i\sqrt{2eBn} I_{n-1,n} & -(\omega +
 k_{3})I_{n-1,n-1} & 0 & mI_{n-1,n-1} \cr }\right ) \nonumber
\ee
and
\be
I_{n,l} = I_{n;k_{2}}(x) I_{l;k_{2}}(x')
\ee
where
\be
I_{n;k_{2}}(x) = \left( {eB \over \pi} \right) ^{1/4} e^{-{1 \over
2}eB\left( x-{k_{2} \over eB} \right) ^{2}} {1 \over 2^{n/2} \sqrt{n!}
} H_{n}\left( \sqrt{eB} \left( x-{k_{2} \over eB}
\right)  \right)    .
\ee
$H_{n}$ are the Hermite polynomials and we have chosen the
$\gamma$-matrices in the chiral representation. The propagator is
obtained by solving Dirac's equation in a magnetic field.  Some useful
identities are
\be
\sum_{n=0}^{\infty}{I_{n;k_{2}}(x)
I_{n;k_{2}}(x')} = \delta (x-x')
\ee
and
\be
\int_{-\infty}^{\infty}{dx I_{n;k_{2}}(x)  I_{l;k_{2}}(x)} =
\delta _{n,l} \label{xint}   .
\ee
In terms of $S(x,x')$ the polarization tensor is
\be
\Pi^{\mu \nu}(x) = e^2 \int {dt'dx'dy'dz'  Tr\left( \gamma ^{\mu}
S(x,x') \gamma ^{\nu} S(x',x) \right) } \label{PIS} .
\ee

Before we pass to the computation of the components of $\Pi^{\mu \nu}$
beyond the static limit we will show how the results of the previous
section can be reproduced using Eq. (\ref{PIS}).

\section{The static limit}

Whenever the momentum insertion from the electromagnetic field into
the fermion loop is vanishing the polarization tensor can be written
\be
\Pi^{\mu \nu} = \sum_{n,l=0}^{\infty}
\int {d\omega dk_{2} dk_{3} \over (2\pi )^{3}}
{1 \over \left( \omega ^{2} -k_{3}^{2}-m^{2}-2eBn \right)
\left( \omega ^{2} -k_{3}^{2}-m^{2}-2eBl \right)}\;\pi ^{\mu \nu}
\ee
where
\be
\pi ^{\mu \nu} = \int dx'~ Tr\left( \gamma ^{\mu} S(n;\omega ,
k_{2},k_{3}) \gamma ^{\nu} S(l;\omega ,k_{2},k_{3}) \right)
\label{pimunu}
\ee
It can be calculated, for example, that
\be
\pi ^{00} =2(\omega ^{2} + k_{3}^{2} +2eBn +m^{2}) \left( I_{n,n} +
 I_{n-1,n-1} \right) \delta _{l,n} \label{pi00f}
\ee
\be
\pi ^{33} =2(\omega ^{2} + k_{3}^{2} -2eBn -m^{2}) \left( I_{n,n} +
 I_{n-1,n-1} \right) \delta _{l,n} \label{piI33}
\ee
\bea
\lefteqn{ \pi ^{11} =2(\omega ^{2} - k_{3}^{2}- m^{2}) (I_{n,n}
\delta _{l,n+1} + I_{n-1,n-1} \delta _{l,n-1}) +}\nonumber \\
&&2eB\left( \sqrt{n(n+1)} I_{n-1,n+1} \delta _{l,n+1} +
\sqrt{n(n-1)} I_{n,n-2} \delta _{l,n-1}\right) \label{pi11f}
\eea
$\pi ^{11}$ differs from $\pi ^{22}$ only in that the last term has
the opposite sign.  As expected $\Pi^{00}$ and $\Pi^{33}$ correspond
to zero angular moment transfer between the photon and the plasma,
implying the selection rule $\Delta n = 0$. Instead, transversally
polarized waves can induce transitions between different Landau
levels.

\subsection{The $T=0$ finite $\mu$ contribution}

Let us consider the $T=0$ contribution for the various components. We
have
\be
\Pi ^{00}(B,0,\mu)  = {8|e|^3 B \over \left( 2\pi  \right) ^{2}}
\sum_{n=\hat{0}}^{\infty} {\oint_C{{d\omega \over 2\pi
i}\int_{0}^{\infty}{dk_3\, {\omega ^{2} +k_3^{2} +2eBn+m^{2} \over
\left(
\omega ^{2} -k_3^{2}-2eBn-m^{2} \right) ^{2}}}}}
\label{pi00mu}
\ee
To obtain this, perform the $k_{2}$ integration using (\ref{xint}).
It follows that $$
\Pi^{00}(B,0,\mu) =
$$ $$ {8|e| ^3B \over \left( 2\pi \right) ^{2}}
\sum_{n=\hat{0}}^{\infty} {\oint{{d\omega \over 2\pi
i}\int_{0}^{\infty}{dk_3\, \left({2\omega ^{2}
\over \left( \omega ^{2} -k_3^{2}-2eBn-m^{2} \right) ^{2}}-{1 \over
\omega ^{2}-k_3^{2}-2eBn-m^{2}}\right) }}} =
$$ $$ {8|e|^3 B \over \left( 2\pi \right) ^{2}}
\sum_{n=\hat{0}}^{\infty}{\left( {d \over
dx}{\int_{0}^{\sqrt{\mu^{2}-m^{2}-x}}{dk_3\sqrt{k_3^{2}+m^{2}+x}}}-\int_{0
}^{\sqrt{\mu^{2}-m^{2}-x}}{{dk_3\, \over 2 \sqrt{k_3^{2}+m^{2}+x}}}
\right)} =
$$
\be
{|e|^3 B \over \pi ^{2}} \sum_{n=\hat{0}}^{\left[ {\mu ^{2}-m^{2}
\over 2eB} \right]}{{\mu \over \sqrt{\mu^{2}-m^{2}-2enB}}} \label{pi00Bmu}  ,
\ee
where $x=2eBn$.  As can be seen from above, the only non-vanishing
contribution is coming from the point where the $\omega = E$ double
pole is crossing the integration contour.  More about the contour $C$
and the methods we use to perform the integrals can be found in the
Appendix.

There are two regions where the expression simplifies.  One is the
limit $B = 0$ where we get
\be
\Pi ^{00}(0,0,\mu) = {e^2 \mu \sqrt{\mu ^{2} -m^{2}} \over 2\pi ^{2}}
\label{pi000mu}
\ee
the other is $2eB> \mu^{2} - m^{2}$ where
\be
\Pi ^{00}(B,0,\mu) = {|e|^3 B \over 2\pi ^{2}} {\mu \over \sqrt{\mu
^{2}- m^{2}}} ~~~.\label{pi00B}
\ee
Eq. ({\ref{pi00B}) can be understood in terms of the modified relation
between the charge density and the chemical potential of the plasma.
In fact, in this limit Eq. (\ref{pairs}) becomes
\be
\rho(\mu)  \approx  {eB\over 2\pi^2} \sqrt{\mu^2 - m^2}
\label{rhomu}
\ee
and Eq.(\ref{pi00B}) can easily be reproduced using $\Pi^{00} = e^2
\partial \rho /\partial \mu$.  Eq. (\ref{pi000mu}) disagrees with the
result of refs.\cite{Midorikawa}.  In fact, the authors obtain zero
for $\Pi ^{00}(0,0,\mu)$.  On the other hand, our Eq.(\ref{pi00Bmu})
agrees completely with the result of ref.\cite{Zeitlin}.

Let us now consider the spatial components and verify that they are
zero. From (\ref{piI33}) we have
\be
\Pi ^{33} (B,0,\mu ) = {8|e|^3 B \over \left( 2\pi  \right) ^{2}}
\sum_{n=\hat{0}}^{\infty} {\oint{
{d\omega \over 2\pi i}\int_{0}^{\infty}{ dk_3~{ \omega ^{2} +k_3^{2}
-2eBn-m^{2} \over \left( \omega ^{2} -k_3^{2}-2eBn-m^{2} \right)
^{2}}}}} = 0
\ee
Again it is important to take the contour crossing into account in
order to obtain the correct result.

$\Pi ^{11}(B,0,\mu)$ and $\Pi ^{22}(B,0,\mu)$ are simply computed.
The last term of (\ref{pi11f}) is killed by the $k_{2}$ integration
and we are left with a vanishing result.
Hence we conclude that
\be
\Pi ^{11}(B,0,\mu) = \Pi ^{22}(B,0,\mu) =  \Pi ^{33}(B,0,\mu) =0
\ee
Thus static magnetic fields are not screened in the present limit.
These equations clearly remain valid when ${\bf B}\rightarrow 0$.
Again we disagree with the conclusion of ref.\cite{Midorikawa} where a
magnetic screening length proportional to the Fermi momenta was
obtained at $B_{\rm ext} = 0$. The disagreement that we obtain here
and for $\Pi^{00}(0,0,\mu)$ suggests to us that the contributions from
the crossing of the poles through the contour $C$ was ignored by the
authors of ref.\cite{Midorikawa}.

Proceeding similarly to what we have done for the other components we
easily deduce that $\Pi ^{0j}(B,0,\mu) =0$ if there is no momentum
insertion in the fermion loop.

\subsection{The finite $T$ contribution}

At finite temperature we must add two more contour integrals:
\bea
\Pi ^{00}(B,T,\mu) &=& {8|e|^3B \over \left( 2\pi  \right)^{2}}
\sum_{n=\hat{0}}^{\infty}  \left(
\int_{-i \infty +\mu -\epsilon}^{i \infty +\mu -\epsilon}
{d\omega \over 2\pi i}~ {1 \over 1+e^{-\beta (\omega -\mu)}}
I(\omega,k,n) \right. \\ &+& \left.\int_{-i \infty +\mu +\epsilon}^{i
\infty+\mu +\epsilon} {d\omega \over 2\pi i}~ {1 \over
1+e^{\beta(\omega -\mu)}} I(\omega,k,n) + \oint_C {d\omega \over 2\pi
i}~ I(\omega,k,n)
\right) \nonumber
\eea
where
\be
I(\omega,k,n) = {\omega ^{2} +k^{2} +2eBn+m^{2} \over
\left( \omega ^{2}-k^{2}-2eBn-m^{2} \right)^2}~~~.
\ee

The integrals are performed according to the prescription given in the
Appendix with the result
\be
\Pi^{00}(B,T,\mu) = {|e|^3 B\beta  \over 4 \pi
^{2}}\sum_{n = \hat{0}}^{\infty}{\int_{0}^{\infty} dk_3~ {\left({1
\over
\cosh ^{2}\left( {\beta (E + \mu) \over 2} \right)}  + {1 \over
\cosh ^{2}\left( {\beta (E - \mu) \over 2} \right)} \right)}}
\ee
in agreement with (\ref{pi00}).  Although here the result coincides
with that of ref.\cite{Midorikawa} in the limit ${\bf B} = 0$, this
can easily be understood since the contributions from the pole
crossing through the contours $C$ and $C_\pm$ cancel in this case.
Similarly one can verify that the spatial components remain zero even
at non-zero temperature.  The behavior of $\Pi^{00}$ as a function of
$B$ is shown in Fig. 1.  As we anticipated in the introduction, we can
see from this figure that the screening properties of the plasma are
sensitive to the magnetic field only when $eB \gg T^2$. Thus the
effect of even a large magnetic field can be neglected if this
condition is not fulfilled.
\begin{center}
\leavevmode
\epsfysize=7cm
\epsfbox{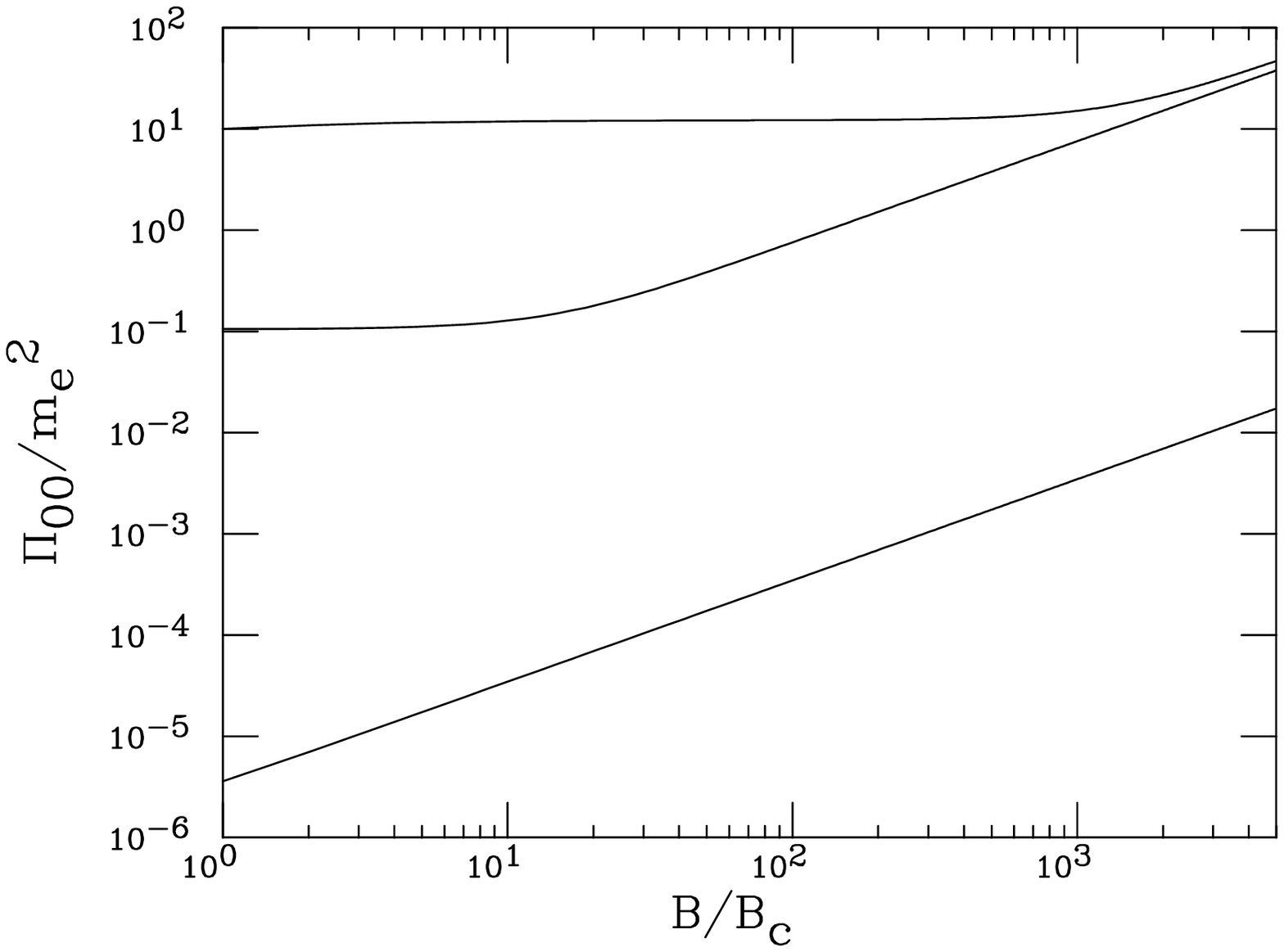}
\end{center}
{\small Fig.1.  $\Pi^{00}$ for three different temperatures. From top,
$T=10m_{e}$, $T=m_{e}$ and $T=0.1m_{e}$.}

\section{Beyond the static limit}

The application of an external magnetic field breaks the isotropy of
the plasma.  It is thus understandable that the electromagnetic
response of the plasma will depend on the direction of the the
momentum insertion with respect to the field orientation. In this
section we will give some sample calculations when the photon has
non-zero momentum or energy.

\subsection{${\bf p} \parallel {\bf B}$}

We start by considering a momentum insertion parallel to the magnetic
field. Put $k'_{3} = k_{3} + p_{3}$ where $p_{3}$ is the momentum
insertion.  $\Pi ^{\mu \nu}$ is then obtained by replacing $k_{3}^{2}$
with $k_{3}'^{2}$ in one of the propagators and replacing $k_{3}^2$
with $k_{3} k'_{3}$ in $\pi^{\mu \nu}$ in
Eqs.(\ref{pi00f}-\ref{pi11f}).  To compute $\Pi^{00}$ we use
\be
I(\omega,k_3,n) = {\omega ^{2} +k_{3}k'_{3} +2enB +m^2 \over \left(
\omega ^{2}   -E_{n}(k_{3})^{2}
 \right)\left( \omega ^{2} -E_{n}(k'_{3})^{2} \right)}
\ee
as the integrand in Eq.(\ref{pi00mu}).  Performing the $\omega$
integration over the contour $C$ (see the appendix) this becomes
\be
{m^2+2eBn \over 2p_{3}}\left( \int_{-\sqrt{\mu ^{2}-m^{2}-2eBn}
}^{\sqrt{\mu ^{2}-m^{2}-2eBn}}{{ dk_3\over
(2k+p_{3})\sqrt{k^{2}+m^{2}+2eBn}}} - (p_{3} \rightarrow -p_{3})
\right) \label{pi00C}
\ee
Expanding to second order in $p_{3}^{2}$ we find $$
\Pi ^{00} (B,0,\mu|p_{3}) = \Pi ^{00} (B,0,\mu|p_{3} =0)  +
$$ $$ {p_{3}^{2}|e|^3B \over \pi ^{2}} \sum_{n=\hat{0}}^{\left[ {\mu
^{2}-m^{2} \over 2eB} \right]} \left( {\mu \over 12 \left( \mu ^{2} -
m^{2} -2eBn \right) ^{3/2}} -{\mu \over 6 \sqrt{\mu ^{2} - m^{2}
-2eBn}(m^2 +2eBn)} \right) $$
\be
 + O(p_{3}^4) \label{pi00py2}
\ee
where $\Pi ^{00}(B,0,\mu)$ is given by Eq.(36).

At $T \neq 0$ we have to manage two simple poles in the complex
$\omega$-plane.  As discussed in the Appendix, the integration over
$C_+$ splits in two integrals each of them corresponding to the two
possible positions of the positive-energy pole: $E > \mu$ or $E <
\mu$. A third integral, over $C_-$, receives a contribution only from
the negative-energy pole, always contained in $C_-$.  An
$T$-independent term coming from the integration over $C_+$ and $C_-$
cancel exactly with the result of the integral over $C$ given by
Eq.(\ref{pi00C}).  Thus we get $$
\Pi ^{00}(B,T,\mu|p_{3}) =
 -{8|e|^3B \over (2\pi)^{2}} \sum_{n=\hat{0}}^{\infty} \frac {2eBn +
 m^2}{p_{3}}\times $$
\be
\int^\infty_{-\infty}  {dk \over (2k + p_{3})\sqrt{k^2+2eBn+m^2}}
 \left( {1 \over 1+e^{\beta (\omega -\mu)}} + {1 \over 1+e^{\beta
 (\omega +\mu)}} \right) ~.
\ee
Previously we found that $\Pi ^{33}(B,T,\mu) =0$ at zero momentum.  At
non-zero $p_{3}$ we have to integrate
\be
{\omega ^{2} +k_{3}k'_{z} -2enB -m^2 \over \left( \omega
^{2}-E_{n}(k_{3})^{2} \right)
\left( \omega ^{2} -E_{n}(k'_{z})^{2} \right)}
\ee
over $k_3$ and $\omega$.  The residue at $\omega = E_{n}$ is
$-\dfrac{k_{3}}{2E_{n}p_{3}}$.  Clearly, $\Pi^{33} = 0$ even at
non-zero $p_{3}$ after integrating over both signs of $k_{3}$. This
agrees with Eq.(\ref{pizz}).

It is also worthwhile to check $\Pi^{11}$ with non-zero $p_3$.  Using
Eq.(\ref{pi11f}) we find that we have now to manage the integrand
\be
{\omega ^{2} -k_3k'_3 - m^{2} \over \left( \omega ^{2}
-k_{3}^2-m^{2}-2neB
\right) \left( \omega ^{2} -k^{'2}_{3}-m^{2}-2(n+1)eBl \right)} \ .
\ee
After the integrations we get
\bea
\lefteqn{\Pi^{11}(B,0,\mu|p_3) = \frac {p_3^2}{B} \left[{|e|
\over 2 \pi ^{2}}
\sum _{n=\hat{0}}^{\infty}  \left( \mu \sqrt{\mu ^{2} -m^{2}-2eBn}
\right.\right.}
\nonumber\\
&& \left.\left. -(m^{2} +4eBn) \log {\mu +
\sqrt{\mu ^{2} -m^{2} -x} \over \sqrt{m^{2}+x}} \right)\right] +
0(p_3^4)
\label{pi113}
\eea
that agree with Eq.(\ref{pixx}) as the reader can check keeping in
mind Eq.(\ref{magn}). The same result as (\ref{pi113}) is found for
$\Pi^{22}(B,0,\mu|p_3)$.

Let us now move on to $\Pi ^{0j}$. Using (\ref{pimunu}) we find that
\be
\pi ^{03} = 2\omega  \left( k_{3} + k_{3}'\right)
\left( I_{n,n} + I_{n-1,n-1} \right) \delta _{n,l}
\ee
This clearly gives zero even for $p_{3} \neq 0$. In fact, the
integrand can be rewritten as
\be
{\omega \over p_{3}}\left( {1 \over \omega ^{2} -k^{2}-2eBn-m^{2}} -
{1 \over \omega ^{2} -k'^{2}-2eBn-m^{2}}\right)
\ee
Shifting the $k'$ in the second term gives zero after integration.  We
leave to verify $\Pi_{01}(B,0,\mu|p_3) = \Pi_{02}(B,0,\mu|p_3) = 0$ as
an easy exercise for the reader.

\subsection{ ${\bf p} \bot {\bf B}$}

The case with momentum transverse to the magnetic field is slightly
more complicated. Of special interest are the $\Pi ^{0i}\ \ (i = 1,2)$
components at non-zero $p_i$ which give through
\be
\sigma ^{ij} = i { \partial \Pi ^{0i} (p) \over \partial p_{j}}
\ee
the plasma conductivity.  Using (\ref{pimunu}) we find that
\be
\pi ^{01} =\int dx'  4i\sqrt{2eBl} \omega \left( I_{n,n} +
I_{n-1,n-1} \right) I_{l,l-1}
\ee
At non-zero $p_{2}$ we have to calculate $$\int dx'dy'dk_{2}dk_{2}'~
e^{ik_{2}(y-y')-ik_{2}'(y-y')+ ip_{2}(y-y')} I_{n;k_{2}}(x)
I_{n';k_{2}}(x')I_{l';k_{2}'} (x')I_{l;k_{2}'}(x) =$$
\be
eB\int dxdx'~ I_{n;0}(x) I_{n';0}(x')I_{l;p_{2}}(x)I_{l';p_{2}}(x')
\ee
This can be found to be
\be
eB e^{-{p_{2}^2 \over 2eB}} \left( -{p_{2} \over(2eB)^{1/2}}
\right)^{n+n'-l-l'} \left( {l!l'! \over n!n'!}\right)^{1/2}
 L_{l}^{n-l} \left( {p_{2}^{2} \over 2eB} \right) L_{l'}^{n'-l'}
\left( {p_{2}^{2} \over 2eB} \right)
\ee
for $n \geq l$ and $n' \geq l'$, where the $L's$ are
Laguerre-polynomials. Expanding to order $p_{2}$ we find, in our case,
\be
 -\sqrt{2eB}p_{2} \left( l \delta _{n,l+1} + l
\delta _{n,l} \right)  + ( n \rightarrow n+1)
\ee
and to order $p_{2}^{3}$ $$ {p_{2}^{3} \over 2\sqrt{2eB}} \left(
\left( 3l^{2} +l \right)
\delta_{n,l+1}
-\left( l^{2}+l \right) \delta _{n,l+2} +\left( 3l^{2}-l \right)
\delta _{n,l} -\left( l^{2}-l \right) \delta _{n,l-1} \right)
$$
\be
+ ( n \rightarrow n+1) .
\ee
For larger powers of $p_{2}$, larger jumps between the Landau levels
are allowed. We get $$
\Pi ^{01} (B,0,\mu |p_{2}) =
{ie^2 p_{2} \over \pi ^{2}} \sum_{n={\hat 0}}^{\left[ { \mu ^{2} -
m^{2}\over 2eB} \right] }\left(
\sqrt{\mu ^{2} -m^{2} -2eBn}
- {eBn \over \sqrt{\mu ^{2} -m^{2} -2eBn} } \right) $$
\be
-{ie p_{2}^{3} \over 4\pi ^{2}B} \sum_{n=1}^{\left[ { \mu ^{2}
- m^{2} \over 2eB} \right]}\left( 2n\sqrt{\mu ^{2}
-m^{2} -2eBn} - {3eBn^{2} \over \sqrt{\mu ^{2} -m^{2}
-2eBn}}
\right) + O(p_{2}^{5})\ . \label{pi01py}
\ee
$\Pi_{02}$ is given by the same expression with $p_2$ replaced by
$p_1$.  Eq.({\ref{pi01py}) confirms the result for the conductivity of
\cite{Zeitlin} and
extends it to second order in momentum. Note that the higher order correction
vanishes for
$2eB > \mu ^{2} -m^{2}$. The expression for the
conductivity given in ref.\cite{Gonzalez}, on the other hand, do not
agree with our result or the result of \cite{Zeitlin}.

We are also able to extend our result to finite temperature.
$$
\Pi ^{01} (B,T,\mu | p_{2}) =
{ie^2 p_{2} \over \pi ^{2}} \sum_{n={\hat 0}}^{\infty}\left( 1 +B{d
 \over dB}\right) \int_{0}^{\infty} dk_3~\left( {1 \over 1+e^{\beta
 (E-\mu )}} - {1 \over 1+e^{\beta (E+\mu )}}
\right)
$$
$$ -{ie p_{2} ^{3} \over 4\pi ^{2}B} \sum_{n=1}^{\infty} \left(
2n +3Bn{d \over dB}\right) \int _{0}^{\infty} dk_3~
\left({1 \over 1+e^{\beta (E-\mu )}} - {1 \over 1+e^{\beta (E+\mu
)}}\right) $$
\be
+ O(p_{2}^{5})
\ee
The first order term can again be checked using (\ref{pi0i}).

We may also consider $\Pi ^{00}$ and $\Pi ^{ii}$ at nonzero $p_{1}$
and $p_{2}$ . The calculation uses the same method as above. After
expanding the Laguerre-polynomials and performing the
contour-integrals, we can verify Eqs.(17-19).  On the other hand we
find
$$
\Pi^{00}(B,0,\mu | p_{2}) = \Pi ^{00} (B,0,\mu, p_2=0) +
$$
$$
{e^2 p_{2}^{2} \over 2\pi ^{2} eB} \sum _{n=\hat{0}} ^{\infty}
\left( \mu \sqrt{\mu ^{2}
-m^{2}-2eBn} + m^{2} \log {\mu + \sqrt{\mu ^{2} -m^{2} -x} \over
\sqrt{m^{2}+2eBn}} -{4eBn\mu \over \sqrt{\mu ^{2} - m^{2} -2eBn}}
\right)
$$
\be
+ O(p_{2}^{4})
\ee
Together with (\ref{pi00py2}) we then have the complete expression to
second order in momentum.  In this case we can not check the result
using section 1.

\subsection{Energy insertion}

Let us now consider the case in which the photon energy is different
from zero. In this case the energy of one of the fermions in the
polarization loop will be shifted to $\omega' = \omega + p_0$ where
the photon energy $p_0$ is assumed to be real.  To compute
$\Pi^{00}(B,0,\mu)$, the integrand to replace the one in Eq.(35) is
\be
I(\omega,k_3,n) = {\omega\omega' +k_{3}^2 +2enB +m^2 \over \left(
\omega ^{2} - E_{n}(k_{3})^{2}
 \right)\left( \omega^{'2} - E_{n}(k_{3})^{2} \right)}~~~.
\ee
Performing the $\omega$-integration we find $$
\Pi^{00}(B,0,\mu|p_0)=
$$ $$ {|e|^3 B \over p_{0} \pi^2}
\left(
\sum_{n = \hat 0}^{\left[ {\mu^{2}-m^{2} \over 2eB} \right]}
\sqrt{\mu^2 - 2enB - m^2} -
\sum_{n = \hat 0}^{\left[ {(\mu -p_{0})^{2}-m^{2} \over 2eB} \right]}
\sqrt{(\mu - p_0)^2 - 2enB - m^2}\right) =
$$ $$
\Pi^{00}(B,0,\mu)+{2|e|^3 B \over \pi^2}
\sum_{n =\hat 0}^{\left[ {\mu^{2}-m^{2} \over 2eB} \right]}\left(
p_0 {2enB + m^2\over 2 (\mu^2 - 2enB - m^2)^{3/2}} + p_0^2 {\mu (2enB
+ m^2)\over (\mu^2 - 2enB -m^2)^{5/2}}\right) $$
\be
+O(p_0^3) ~~~.
\ee
At $T\neq 0$ we have instead $$\Pi^{00}(B,T,\mu|p_0) = {8|e|^3 B
\over(2\pi)^2 p_{0}}\sum ^{\infty}_{n=\hat 0}
\left[\int^{\infty}_{-\infty} dk_3~\left(\frac 1
{e^{\beta(E-\mu)}+1} - \frac 1 {e^{\beta(E+\mu)}+1}\right)\right.  $$
\be
-\left.\int^{\infty}_{-\infty} dk_3~\left(\frac 1
{e^{\beta(E-(\mu+p_0))}+1} - \frac 1 {e^{\beta(E+(\mu+p_0))}+1}
\right)\right]
\ee

\section{Conclusions}

In this paper we have studied the polarization tensor of a QED plasma
placed in a magnetic field.  Although this has been the subject of
several studies, some discrepancies, even in the simplest limit $B =
0$ \cite{Midorikawa}, called for a more careful analysis.
Furthermore, recent results concerning the thermodynamics of a QED
plasma in a strong magnetic field allow a new, physically more
transparent, interpretation of old results as well as of the new
results that we obtain here.

We have used two different methods.  The first relate the polarization
tensor directly to the thermodynamic quantities of the system. The
second makes use of the fermion propagator in an external magnetic
field. These methods give the same results in the first non-vanishing
order in the photon 4-momentum.  Beyond this limit the first method is
not applicable and the introduction of the second can not be avoided.
For instance, this is needed if one wants to find the dispersion
relations for the electromagnetic waves.

Some of our results have a simple physical interpretation.  Static
electric fields are screened by the plasma in strong magnetic fields
($eB > T^2, \mu^2 - m^2$) more effectively than in the free field
case.  This can be understood since static electric fields are
screened by the charge rearrangement in the plasma. At $T = 0$ no
thermal pair production is active and the screening can be achieved
only in the presence of a charge asymmetry. In strong magnetic fields
the relation with the charge asymmetry is modified. Indeed, the charge
density becomes proportional to the magnetic induction when $eB$ is
larger than the Fermi momentum squared, $\mu^2 - m^2$ (see
Eq.(\ref{rhomu})). Consequently, $\Pi^{00}$ grows linearly with $B$
(see Eq.(\ref{pi00B}).

At $T\neq 0$, as we discussed in the introduction, the thermal pair
production is amplified by large magnetic fields. Then more charge
carriers are available to screen static electric fields. The reader
can see in Fig. 1 that $\Pi^{00}$ starts to grow linearly in $B$, once
all electrons and positrons have dropped in the lowest Landau level.
In this static limit the anisotropy induced by the external magnetic
field can not play any role and the definition of an electric
screening length $\lambda^{-2} = m_{el}^2 = \Pi^{00}(p_0=0, {\bf p}
\rightarrow 0)$
\cite{Kapusta} is still meaningful.

Static magnetic fields are not screened at all. In fact, macroscopic
spatial currents can not be obtained if ${\bf p} = 0$ (see
Eq.\ref{piij}). This conclusion can not be modified in presence of
strong external magnetic fields neither at vanishing nor at finite
temperature.  However, magnetic screening is achieved if ${\bf p} \neq
0$ and Eqs.(\ref{pizz}-\ref{piyz}) show how this is related to the
magnetization and magnetic susceptibility of the plasma in a non
trivial way.

We also computed the electric conductivity of the plasma. Our result
confirms that of ref.\cite{Zeitlin}, but we improve the calculation by
including second order terms in the momentum expansion. On the other
hand, we disagree with the result of ref.\cite{Gonzalez}.

The determination of the complete dispersion relations for
electromagnetic waves propagating trough plasmas in strong magnetic
fields is beyond the purposes of the present paper. One would need to
take into account the full tensorial structure of the polarization
operator as done in Refs.\cite{Rojas}.  Using our results according to
the prescription of Refs.\cite{Rojas}, would provide physically more
accessible information about the propagation of electromagnetic waves
trough a plasma in presence of strong magnetic fields.

\section*{Acknowledgements}

We would like to thank Lars Bergstr\"om and Hector Rubinstein for
discussions and encouragement while writing this paper. The work of
D. G. was supported by a Twinning EEC contract.

\section*{Appendix}

Let us give some details on how the contours are specified. At finite
temperature the integral $\int d\omega$ is replaced by a sum over
$\omega ={i\pi \over \beta} (2n+1)$. This can be accomplished by
integration along a counter clockwise contour around the poles of
$\tanh{{\pi z \over 2}}$, e.g.
\be
\sum _{n} f(2n+1) = {i \over 4 } \oint dz \tanh {\pi z \over 2} f(z)
\ee
Deforming the contours gives
\bea
\lefteqn{ \int _{-i\infty}^{i\infty} \frac {d\omega}{2\pi i} f(\omega )
\int_{-i \infty +\mu +\epsilon}^{i \infty +\mu +\epsilon}
\frac {d\omega}{2\pi i}   f(\omega)
{1 \over e^{\beta (\omega -\mu)}+1} +} \\ && \int_{-i \infty +\mu
-\epsilon} ^{i \infty +\mu -\epsilon}
\frac {d\omega}{2\pi i}   f(\omega)
{1 \over e^{-\beta (\omega -\mu)}+1} + \oint_C
\frac {d\omega}{2\pi i}  f(\omega)
\nonumber
\eea
The first term is the divergent vacuum contribution, the second and
third terms are non-zero only at finite temperature and the last term
is the $T=0$ non-zero $\mu$ contribution.  The contours are depicted
in fig. 2.

\begin{center}
\leavevmode
\epsfysize=7cm
\epsfbox{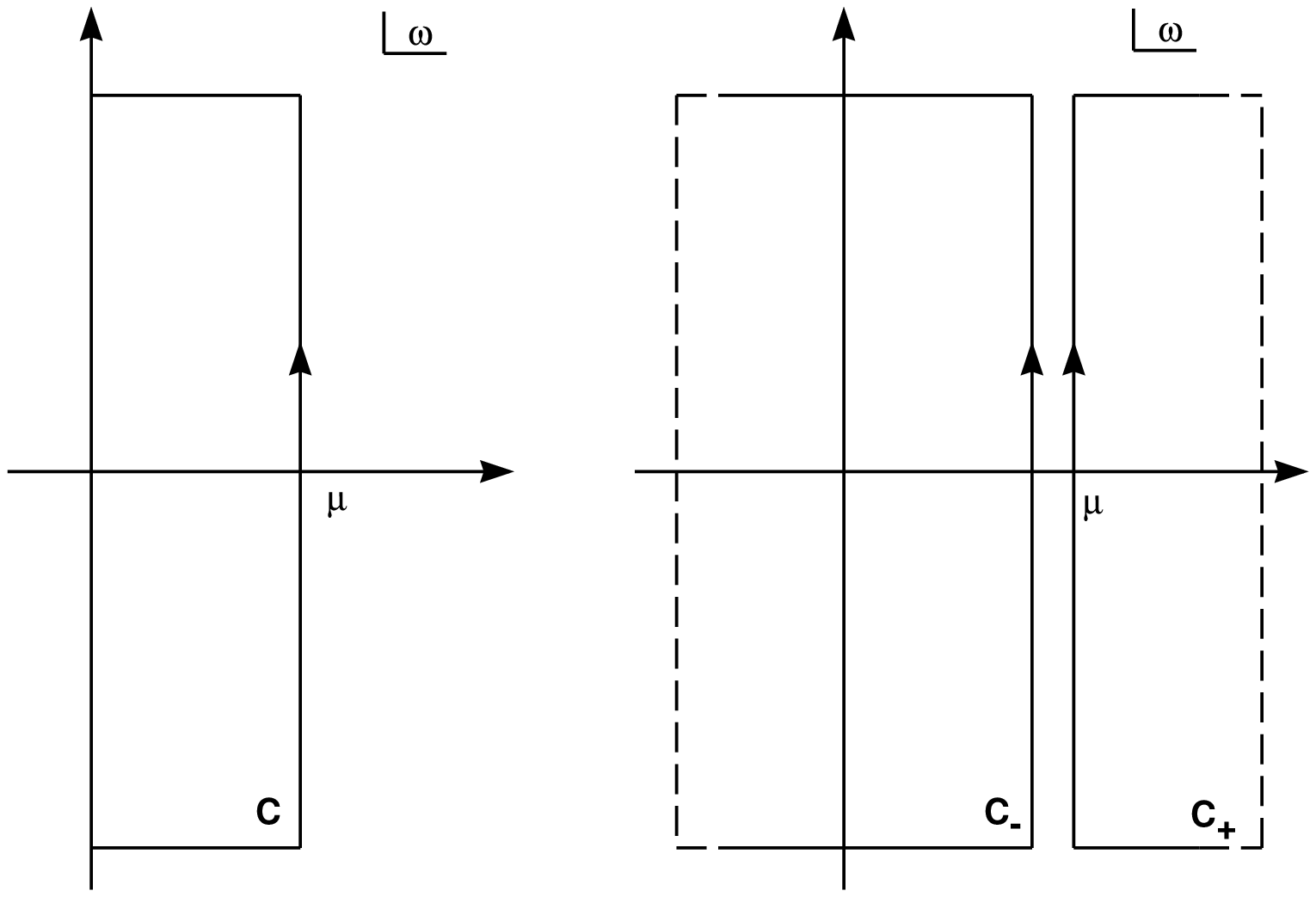}
\end{center}
{\small Fig. 2. Contours for $T = 0$ on the left and for $T \neq 0$ on
the right.

\vskip 1cm

It is worthwhile also to give some details about how we performed the
integrals. Mainly we wish to direct the attention of the reader to the
crossing of poles through the contours of integration.

We start by considering the $T = 0$ finite-density contribution.  This
term has the form
\be
\oint_C \frac {d\omega}{2\pi i}\int^{\infty}_{-\infty} dk
f(\omega,k) = \sum_{i}{\rm lim}_{\omega \rightarrow - E_i}
\frac {d^{m_i-1}}{d\omega^{m_i-1}}\int^{\infty}_{-\infty} dk
f(\omega,k) (\omega - E_i)^m_i\theta(\mu - \omega)
\label{intC}
\ee
where we have assumed that the function $f(\omega,k)$ has poles
$\omega = E_i$ on the real axis of order $m_i$.  The contour of
integration $C$ is depicted in Fig. 2.  The result of the integration
is zero if the poles lie outside the region delimited by the contour
$C$. This justifies the $\theta$-function on the left side of
Eq.(\ref{intC}).  The effect of the $\theta$-function is simply to
modify the effective integration limits in the case the pole is
first-order.

More interesting, when the pole is second-order (we do not come
through higher order poles in our one-loop computations) a new term
appears containing a delta-function coming from the derivative of the
$\theta$. This term corresponds to the crossing of the pole trough the
contour of integration. The $\delta$ kills the $k$-integral and leaves
us with a term of the form like that in the result of Eq.(36).

The finite temperature contribution has the general form
\bea
\lefteqn{\int_{-i \infty +\mu +\epsilon}^{i \infty +\mu +\epsilon}
\frac {d\omega}{2\pi i}  \int_{-\infty}^{\infty} dk~ f(\omega,k)
{1 \over e^{\beta (\omega -\mu)}+1} +} \\ && \int_{-i \infty +\mu
-\epsilon} ^{i \infty +\mu -\epsilon}
\frac {d\omega}{2\pi i}  \int_{-\infty}^{\infty} dk~ f(\omega,k)
{1 \over e^{-\beta (\omega -\mu)}+1} + \oint_C
\frac {d\omega}{2\pi i} \int_{-\infty}^{\infty} dk~ f(\omega,k)
\nonumber \label{Tint}
\eea

When the poles are first order the first two, temperature dependent,
integrals split in three non-vanishing parts. The first part
corresponds to the integral over the contour $C_+$ (see Fig. 2.) for
values of $k$ such that the pole $\omega = E$ is inside $C_+$ ($E >
\mu$). It can be written
\be
\int^{\infty}_{-\infty} dk~ f(E,k) \frac 1 {e^{\beta(E - \mu)}+1} -
\int^{\sqrt{\mu^2 - x - m^2}}_{-\sqrt{\mu^2 - x - m^2}} dk~ f(E,k)
\frac 1 {e^{\beta(E - \mu)}+1}~~~; \label{C+}
\ee
the second part corresponds to the integral over $C_-$ when only the
pole $\omega = -E$ is inside $C_-$
\be
\int^{\infty}_{-\infty} dk f(-E,k) \frac 1 {e^{\beta(E + \mu)}+1}
\ee
and the third one comes from the contribution to the same integral
when $k$ is such that the pole $\omega = E$ is inside $C_-$
\be
\int^{\sqrt{\mu^2 - x - m^2}}_{-\sqrt{\mu^2 - x - m^2}} dk~ f(E,k)
\frac 1 {e^{-\beta(E - \mu)}+1}~~~.\label{C-2}
\ee
It is then straightforward to verify that the sum of the second
integral in (\ref{C+}) and the integral in (\ref{C-2}) is temperature
independent and cancels with the integral over $C$ in
(\ref{Tint}). Then (\ref{Tint}) takes the form
\be
\int^{\infty}_{-\infty} dk~ f(E,k) \left(\frac 1 {e^{\beta(E - \mu)}
 + 1} + \frac 1 {e^{\beta(E + \mu)} + 1}\right)
\ee
where the separate contributions of electrons and positrons are
evident.  If the poles are second-order we have to take care of the
contribution coming from the derivative of the $\theta$-function
integrated through the $C_{\pm}$ contours. This generates the terms $$
\int^{\infty}_{-\infty} dk~ f(E,k)
\frac 1 {e^{\beta(E - \mu)} + 1} \delta(E - \mu)
+ \int^{\infty}_{-\infty} dk~ f(E,k)
\frac 1 {e^{-\beta(E - \mu)} +1} \delta(E - \mu) =
$$
\be
\frac 1 2  \left(\frac {dE}{dk}\right)^{-1}_{E = \mu}
f(E = \mu) = \frac \mu {\sqrt{\mu^2 - x - m^2}}~f(E = \mu) ~~~.
\ee
This is a T independent term. When inserted in (\ref{Tint}) it cancel
the result of the integral over $C$.

\end{document}